\documentclass[aps,prl,twocolumn,showpacs,groupedaddress]{revtex4}
\usepackage{graphicx}

\newlength\figurewidth
\begin{document}
\newcommand{\rem}[1]{}

\title{ Ferromagnetically coupled magnetic impurities in a quantum
point contact }

\author{Taegeun Song}
\author{Kang-Hun Ahn}
\affiliation{Department of Physics, Chungnam National University,
 Daejeon 305-764, Republic of Korea}

\date{\today}

\begin{abstract}

We investigate the ground and excited states of interacting
electrons in a quantum point contact using exact diagonalization
method. We find that strongly localized states in the point contact
appear when a new conductance channel opens due to momentum
mismatch. These localized states form magnetic impurity states which
are stable in a finite regime of chemical potential and excitation
energy. Interestingly, these magnetic impurities have ferromagnetic
coupling, which shed light on the experimentally observed puzzling
coexistence of Kondo correlation and spin filtering in a quantum
point contact.
\end{abstract}


\maketitle

 Since two decades ago, it has been known
that the conductance of a quantum point contact(QPC), a narrow
constriction between two-dimensional electron gas, has plateaus at
integer multiples of $2e^{2}/h$\cite{wees}. This conductance
quantization is due to the fact that the density of states of
one-dimensional conductor
 is inversely proportional to the electron velocity, which is now well understood within Landauer
formalism\cite{mesoscopics}. Additional structure of the extra
plateaus around $0.7 (2e^{2}/h)$,
 has been observed, but the origin of the structure has been controversial ever since its
observation\cite{thomas,cronenwett,roche,rokhinson}. It has been
suggested that the anomalous feature might be due to Kondo
correlation because, similar to Kondo effect, there exist a
zero-bias peak in the differential conductance which splits in a
magnetic field and a crossover to perfect transmission below a
characteristic temperature\cite{cronenwett}.

The Kondo interpretation as an origin of 0.7 structure, however, is
questioned mainly by two reasons. First, the counter-intuitive
existence of the impurity state on top of the potential barrier.
Second, spin-filtering effect has been observed in a quantum point
contact\cite{rokhinson,thomas}, which is hardly explainable within
the Kondo model. The spin filtering effect is better explained by a
phenomenological spin polarization model\cite{reilly} which assumes
also counter-intuitive low-dimensional spontaneous spin
polarization. In this Letter, we provide generalized view allowing
both Kondo correlation and spin filtering by confirming the
existence of the ferromagnetically coupled magnetic impurities in a
quantum point contact. We will demonstrate the existence of the
magnetic impurity in a QPC through exact diagonalization technique
and show that the magnetic impurities are ferromagnetically coupled.
Our numerical results imply that the transport through a QPC may
have both the Kondo correlation and spin-polarised transport due to
the interplay between ferromagnetically coupled magnetic impurities.

\begin{figure} \includegraphics[width=.5\textwidth]{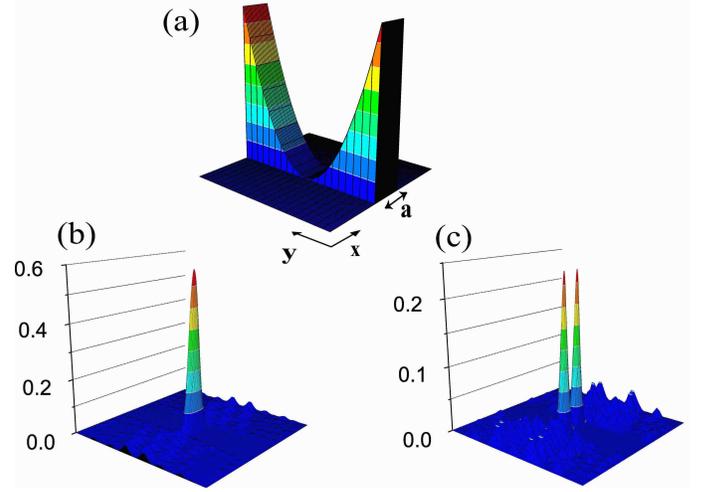}
 \caption{
(a) Three dimensional plot of electric potential for a quantum point
contact.
 (b) Probability density $| \psi_i (\bf{r})|^2$ of 206th eigen-state showing the first
 resonant level in QPC, and (c) the probability density of the 238th
 state showing the second resonant level. The parameters for the
 system size
 $L_x=L_y=43.30 l_{0}$  and $a=8.66 l_{0}$  where
 $l_0=\sqrt{\frac{\hbar}{m^{*}\omega_0}}$.
 }
\end{figure}

Let us consider a quantum point contact modeled by a harmonic
potential locally formed in a two dimensional electron system shown
in Fig. 1 (a). The single particle Hamiltonian $h$ for this system
is given by \begin{eqnarray}
h&=&-\frac{\hbar^{2}}{2m^{*}}(\frac{\partial^{2}}{\partial
x^{2}}+\frac{\partial^{2}}{\partial y^{2}})+v(x,y)\\
v(x,y)&=&\frac{1}{2}m^{*}\omega_{0}^{2}y^{2}\Theta(\frac{a}{2}-x)\Theta(\frac{a}{2}+x),
\label{shamilt} \end{eqnarray} where $m^{*}$ is the band effective
mass of electron, $\omega_{0}$ is the natural angular frequnecy of
the harmonic confinement of the point contact in transverse
direction, and $\Theta(x)$ is the Heaviside step function. We impose
periodic boundary condition for x direction ($-L_{x}/2<x<L_{x}/2$)
and hard-wall boundary condition for y direction
($-L_{y}/2<y<L_{y}/2$). Here, $|x|<a/2$ is the regime of the quantum
point contact and $a/2<|x|<L_{x}/2$ describes the regime of the
electron reservoir. By diagonalizing the Hamiltonian in
Eq.(\ref{shamilt}) using eigen-wavefunction for $v=0$ as basis wave
functions, we obtain eigen-wavefunction $\psi_{i}({\bf r})$
satisfying $h\psi_{i} = \epsilon_{i}\psi_{i}$.

When a new transport channel opens near
$\epsilon=(n+\frac{1}{2})\hbar\omega_{0}$, transport states which
have the electron velocity $\frac{1}{\hbar}\frac{d\epsilon}{dk}$
contribute to the electric conductance. The conductance quantization
is due to the fact that the density of the transport states,
 $\frac{1}{2\pi}\frac{dk}{d\epsilon}$, is proportional to the inverse of the velocity.
This is Landauer formalism\cite{mesoscopics}, where it is assumed
that an electron out of the conductor does not scatter back to the
conductor.
 The back-scattering, however, becomes significant
when the momentum mismatch at the ends of the conductor is serious.
In terms of the parameters for the QPC model in Eq.(\ref{shamilt}),
we can write the momentum mismatch condition can be written as
 $\hbar\omega_{0}
\gg \frac{\pi^{2}\hbar^{2}}{m^{*}a^{2}}$. If this condition is met,
the wavefunction at the ends of the point contact, $x=\pm
\frac{a}{2}$, is close to zero due to the continuity of
$\frac{\partial \psi}{\partial x}$.

\begin{figure}
\includegraphics[width=.45\textwidth]{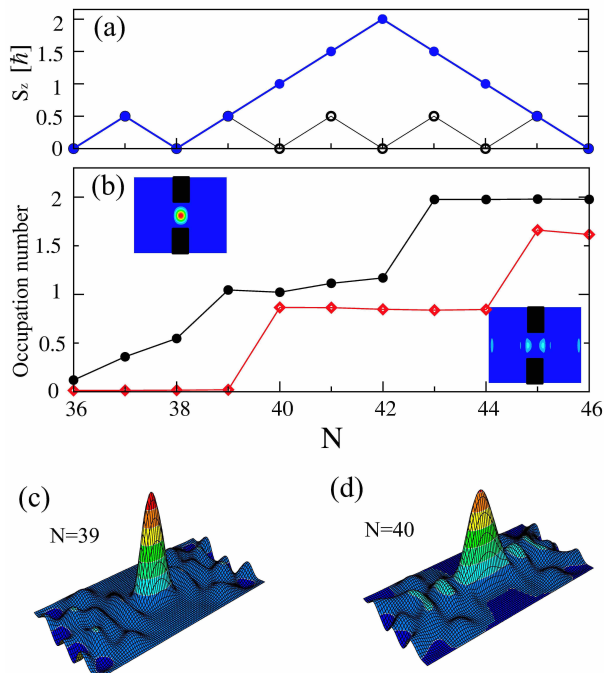}
 \caption{(color online)
(a) The net spin of the ground state of interacting
 electrons as a function of the total number of electrons N ( filled circle )
 compared to the net spin of non-interacting electrons (empty circles).
  (b) The
 occupation number of the first (second) resonant level where its probability density is shown in the left(right)  inset denoted
 by filled  circles (empty diamonds). The lines are guide to the eye.
  The total spin density of the ground state of 39 electrons
 (c) and 40 electrons (d). We used parameters of $L_x=23.95 l_0, L_y=10.79 l_0$, $a=5.06
 l_0$ and interaction strength $l_{0}=0.32a_B^*$ where $a^{*}_{B}$ is the effective Bohr radius.)
 }
\end{figure}

The resonance condition for the existence of the standing wave in
QPC, $ka=\pi,2\pi,3\pi,\cdots$, gives us good estimation of the
electron energies showing the resonance;
 \begin{eqnarray}
\epsilon_{nl} \approx (n+\frac{1}{2})\hbar\omega_{0}
+\frac{\pi^{2}\hbar^{2}l^{2}}{2m^{*}a^{2}},~~~(\hbar\omega_{0} \gg
\frac{\pi^{2}\hbar^{2}}{m^{*}a^{2}} )
\label{dispersion}
 \end{eqnarray}
 where $n=0,1,2,\cdots$ is the transport channel index and $l=1,2,3,\cdots$
 is the index for the resonant levels.
Our numerical results confirm the estimation in
Eq.(\ref{dispersion}) (not shown).
 Interestingly, low-level resonance shows
strong enhancement of probability density $|\psi|^{2}$ in the QPC
regime (See Fig. 1 (b) and (c) ). Note that, in contrast to the
previous analysis\cite{rejec}, the localization here does not
originate from electron interactions but from momentum mismatch. As
will be shown later, due to the electron interaction, the localized
states become singly occupied and thus magnetic impurities.

An Anderson magnetic impurity model has been introduced to account
for the Kondo features of 0.7 anomaly\cite{meir}.
 Numerical
calculations based on spin-density functional theory are suggestive
of the existence of the magnetic impurity states\cite{rejec,hirose}.
Since the density functional theories, however, are valid only for
the ground state, it is questionable whether spin density functional
theory can be applied to the Kondo effect where the magnetic
impurity plays dynamical role. The exact diagonalization method can
be used to investigate not only the ground state but also excited
states of interacting electrons, thus it allow us to study whether
the magnetic impurity can survive low energy excitation.

To obtain eigenstates of interacting electrons, we use a many-body
Hamiltonian for electrons in the form;
\begin{eqnarray}
H=\sum_{i,\sigma}\epsilon_{i,\sigma}c^{\dagger}_{i,\sigma}c_{i,\sigma}
+\frac{1}{2}\sum_{i,j,k,l,\sigma,\sigma^{\prime}}V_{ijkl}c^{\dagger}_{i,\sigma}c^{\dagger}_{j,\sigma^{\prime}}
c_{l,\sigma^{\prime}}c_{k,\sigma},
\end{eqnarray}
where $c_{i,\sigma}$ is annihiliation operator for the i-th single
particle eigenstate of h with spin $\sigma$, and
\begin{eqnarray}
V_{ijkl}=\int d{\bf r}\int d{\bf r^{\prime}}\psi_{i}^{*}({\bf
r})\psi_{j}^{*}({\bf r^{\prime}})\frac{e^{2}}{|{\bf r}-{\bf
r^{\prime}}|}\psi_{k}({\bf r})\psi_{l}({\bf r^{\prime}}).
\end{eqnarray}
We use Slater-determinant states for spin-up and spin-down electrons
such as $|\bullet \circ \circ \bullet \bullet \circ \cdots | \bullet
\bullet \circ \circ \bullet \cdots >$ for many-electron base
vectors, where $\bullet$ and $\circ$ denote the occupied and
unoccupied state, respectively. The circles in the left (right) of
$|$ denote the spin up (down) states. We calculate many electron
eigenstates up to 50  electrons where maximum matrix size is up to
$11532 \times 11532$.

 In Fig. 2 (a), we plot the net spin in units of $\hbar$ ($S_{z}=\frac{1}{2}(N_{\rm up}-N_{\rm down})$) of
the ground state as a function of the total number of electrons
$N=N_{\rm up}+N_{\rm down}$. In usual set-up of experiments, a gate
voltage $V_{g}$ of quantum point contact is proportional to the
chemical potential $\mu$. The total number of electrons $N$ is
roughly proportional to the gate voltage since ;
\begin{eqnarray}
V_{g} \propto \mu =E_{N}-E_{N-1} \propto N,
\end{eqnarray}
 where $E_{N} \propto N^{2}$ is the
ground state energy of $N$ interacting electrons. Thus, the gate
voltage $V_{g}$ dependence of the spin polarization can be probed by
varying the number of electron $N$ as in Fig.2 (a).
 The
electron-electron interaction induces spin polarization of the
  electrons in the ground state in consistent with
  the density functional theory calculations\cite{rejec}.
The spontaneous spin polarization is due to Hund coupling of
electrons which are strongly bounded in the quantum point contact.
Fig. 2 (a) and (b) show that the appearance of the spin-polarized
regime ($39\le N \le 45$) accompanies the single occupancy of the
resonant levels. This is the regime where a transport channel begins
to open as the gate voltage increases.

\begin{figure} \includegraphics[width=.5\textwidth]{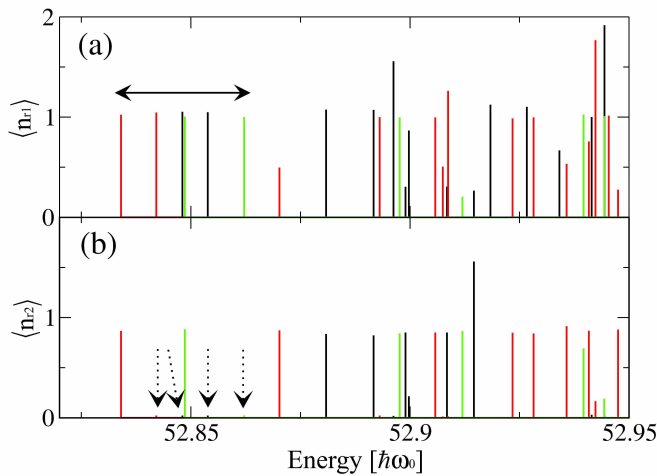}
 \caption{ (colors on line) The
occupation number of the resonant level r1 (a) and r2 (b) for 40
electrons.
 we used the same parameters used for Fig. 2. The net spin of
 $S_{z}=0, 1, 2$ of each state is denoted by the color of black,
 red, and green.
 }
\end{figure}

The single occupancy $<n_{i}>=<n_{i\uparrow}>+<n_{i\downarrow}>
\approx 1$ is necessary condition for the magnetic impurity in the
Kondo effect. This indeed happens as shown in Fig. 2 (b). As the
total number of electron $N$ (or the gate voltage ) increases, the
first resonant level ( say $r1$, of which  probability density shown
in the left inset) becomes singly occupied ( filled circle ). There
exists a finite regime of $N$ ( $39\le N \le 42$) and so the
corresponding regime of $V_{g}$ for $<n_{r1}>\approx 1$. The first
resonant level becomes no more magnetic impurity as it becomes
doubly occupied for $N \ge 43$. Instead, the second resonant level,
$r2$, takes the role of magnetic impurity for $N\ge 43$ (empty
diamonds). In the intermediate regime ($40\le N \le 42$), both of
the resonant levels $r1$ and $r2$ are singly occupied.

The polarized spins are distributed mainly near the center of
quantum point contact as shown in Fig. 2 (c) and (d) where we plot
the spin density $<S_{z}({\bf r})>$ for the ground states of $N=39$
and $N=40$;
 \begin{eqnarray}
 S_{z}({\bf r})=\sum_{i,j}\psi^{*}_{i}({\bf
r})\psi_{j}({\bf r})
[c_{i\uparrow}^{\dagger}c_{j\uparrow}-c_{i\downarrow}^{\dagger}c_{j\downarrow}
]
 \end{eqnarray}
Note that by adding one more electron to $S_{z}=1$ state (Fig. 2 (c)
), the spin in the regime of quantum point contact is enhanced to be
$S_{z}=2$ (Fig. 2 (d)) , i.e., the spins of singly occupied states
are {\it ferromagnetically}  coupled.

  In Fig. 3, we plot the occupation numbers
of two localized states $r1$ and $r2$ where their probability
densities are shown in the left and right inset of Fig. 2 (b),
respectively. If the magnetic impurity picture is valid for a
quantum point contact at finite bias and temperatures, there must be
low-lying many-electron excited states which have single occupancy
in the localized state. This is indeed the case for the first
resonant level $r1$ so that $<n_{r1}>\approx 1$ denoted by the
horizontal solid arrow in Fig. 3 (a). However, the second resonant
level is singly occupied $<n_{r2}>\approx 1$ only in the ground
state and it becomes almost empty even in the first excited state
$<n_{r2}>\approx 0$ as shown by the dotted arrow in Fig. 3 (b). This
implies that the localization in the second resonant level is
unstable against energy excitation. The electron in the second
resonant level can easily escape out of QPC by applying a weak bias
when the first resonant level is singly occupied, which could not be
seen in the previous study based on density functional theory.
 When the gate voltage increases further, the second level again forms a stable magnetic impurity and the first
 level becomes doubly occupied ( not shown).

The observation mentioned above allows us to introduce an Anderson
Hamiltonian for two resonant levels which describes low-lying
excitations of the interacting electrons in QPC,
\begin{eqnarray} \nonumber
H&=&\sum_{\sigma;i=1,2}\epsilon_{i\sigma}d_{i\sigma}^{\dagger}d_{i\sigma}+
 \sum_{\sigma;k\in
L,R}\epsilon_{k\sigma}c_{k\sigma}^{\dagger}c_{k\sigma}\\ \nonumber
&+&\sum_{i=1,2}U^{(i)}n_{i\uparrow}n_{i\downarrow} +
\sum_{\sigma;i=1,2;k\in L,R}
[V_{k\sigma}^{(i)}c_{k\sigma}^{\dagger}d_{i\sigma} +H.c.] \\
 &-&J_{F}\vec{S}_{1}\cdot \vec{ S}_{2}, \label{model}
\end{eqnarray} where $n_{i\sigma}=d^{\dagger}_{i\sigma}d_{i\sigma}$,
$\vec{
S}_{i}=\sum_{\sigma,\sigma^{\prime}}d_{i\sigma}^{\dagger}{\vec{
\sigma}}_{\sigma,\sigma^{\prime}}d_{i\sigma^{\prime}}$,
$\epsilon_{1\sigma}<\epsilon_{2\sigma}$ are on-site energies of the
localized states, $U^{(1)} >U^{(2)}$ are their on-site Coulomb
energies, $J_{F}$ is the ferromagnetic coupling strength between the
localized states. $V^{(1)}_{k\sigma}<V^{(2)}_{k\sigma}$ is the
coupling strength between the localized state and the lead states
created by $d^{\dagger}_{i,\sigma}$ and $c^{\dagger}_{k,\sigma}$,
respectively.

While the conductance calculation of the above model is not
performed here, we sketch the possible transport mechanisms expected
from the model in Eq.(\ref{model}) ( See Fig. 4). When the chemical
potential $\mu$ approaches the value for a new channel opening,
$\epsilon_{1} < \mu <\epsilon_{2}-J_{F}$, the first localized state
becomes a magnetic impurity giving rise to the Kondo transport (see
Fig. 4 (a)).
 If both of the localized states are
singly  occupied when $\epsilon_{2}-J_{F}< \mu <
\epsilon_{1}+U^{(1)}$, Kondo resonance through the first localized
state will be hindered due to the ferromagnetic coupling with the
second localized state. In this regime, spin-polarized transport is
possible via enhanced g-factor through ferromagnetic coupling
between the localized states. The spin polarized transport will be
again weakened, if the first localized state becomes doubly occupied
when $\mu
> \epsilon_{1}+U^{(1)}$.

 In the experiments showing spin filtering effect\cite{rokhinson,thomas}, while an external magnetic field was applied,
its strength was considered to be too small to polarize electron
spin ($g\mu_{B}B < k_{B}T$). The ferromagnetic coupling between the
localized states, however, can strongly enhance the spin splitting.
The ferromagnetic coupling can enhance the spin splitting of the
second localized state $ \Delta \approx g\mu_{B}B + 2J_{F}(<n_{\rm
1\uparrow}>-<n_{\rm 1\downarrow}>) \approx
g\mu_{B}B+2J_{F}\tanh(\frac{g\mu_{B}B}{2k_{B}T})$. Provided the
ferromagnetic coupling $J_{F}$ is large enough compared to the
thermal energy, the spin splitting of the first localized state is
given by $\Delta \approx J_{F}\frac{g\mu_{B}B}{k_{B}T} ~~~(J_{F}\gg
k_{B}T \gg g \mu_{B} B).$

The condition for the spin-polarized transport, $\Delta > k_{B}T$,
then becomes a experimentally reachable condition;
\begin{eqnarray}
\sqrt{J_{F}g\mu_{B}B} > k_{B}T.
\end{eqnarray}
In the spin polarization measurement
experiments\cite{rokhinson,thomas}, typical electron interaction
energy and the ferromagnetic coupling is about the order of
$J_{F}\sim$ 1 meV, the bare Zeeman splitting is the order of
$g\mu_{B}B \sim 1 \mu$eV. Thus, the spin splitting in a quantum
point contact is expected to be enhanced by two orders of magnitude,
$\sqrt{\frac{J_{F}}{g\mu_{B}B}}\sim 100$, which makes the effective
spin splitting $\Delta$ comparable to the thermal energy $k_{B}T$.
It is worth noting that similar models on the transport through
double dot show spin-polarized transport\cite{vernek,lipinski}.

\begin{figure}
\includegraphics[width=.4\textwidth]{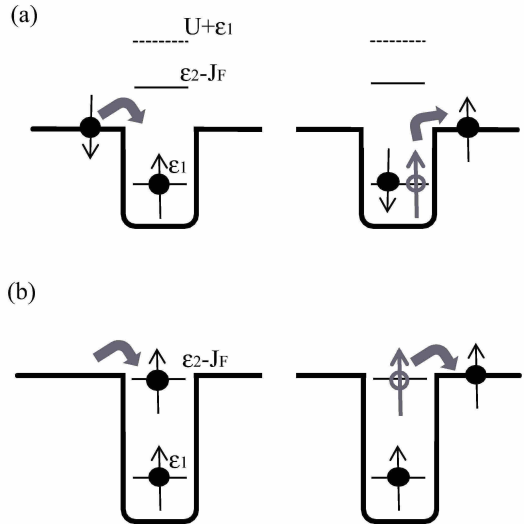}
 \caption{ (a) Schematic figure for the Kondo type transport through the
 first resonant level and (b) spin polarized transport through the
 second localized level.
 }
\end{figure}

In summary, our numerical results based on exact diagonalization
method show that localized states appear as resonant energy levels
when a new conductance channel opens, and the magnetic impurities
indeed exist as also excited states of interacting electrons. Thus,
we confirm the Kondo magnetic impurity in a quantum point contact.
Interestingly, the magnetic impurities have ferromagnetic coupling
with each others which is expected to cause spin polarized
transport. Our numerical calculation strongly supports that the
counter-intuitive coexistence of Kondo correlation and spin
filtering in a quantum point contact is due to ferromagnetically
coupled magnetic impurities in the quantum point contact. The
existence of the magnetic impurity in our work is in consistent with
the recent experimental evidence for the bound state in quantum
point contact probed with coupled QPCs\cite{yoon}.

We thank B. Wu, J. Hong, K. Richter, Y. Nazarov for fruitful
discussion.
 We acknowledge Max Planck Institute for
Physics of Complex Systems for allowing us to use parallel computing
facility and hospitality during the authors' visit.
 This work was supported by the National Research Foundation funded
by the Korea government (No.KRF-2008-C00140).

\vspace{-0.7cm}

\bibliography{apssamp}


\end{document}